\documentclass[12pt,a4paper]{article}
\usepackage{graphicx}
\usepackage{graphics}
\usepackage{epsfig,multicol}
\usepackage{amsmath}
\usepackage{bm}
\usepackage{multirow}
\usepackage{color}
\usepackage{hyperref}
\newcommand{\tp}{${t^{\prime}}$}

\newcommand{\TP}{${\bar{t}^{\prime}}$}
\begin{document}
\begin{flushright}
RECAPP-HRI-2012-003
\end{flushright}
\begin{center}
{\Large\bf 
A clean signal for a top-like isosinglet fermion at the Large hadron Collider}\\[20mm]
Aarti Girdhar{\footnote{aarti@hri.res.in}}~ and~ Biswarup Mukhopadhyaya
{\footnote{biswarup@hri.res.in}} \\
{\it Regional Centre for Accelerator-based particle Physics}\\
{\it Harish-Chandra Research Institute}\\ 
{\it Chatnaag Road, Jhunsi, Allahabad, 211019,India} 
\end{center}
\begin{abstract}
We predict a clean signal at the Large Hadron Collider ($\sqrt{s}$=14TeV) for 
a scenario where there is a top-like, charge $+ \frac{2}{3}$ 
vectorlike isosinglet fermion. Such a quark, via mixing with the
standard model top, can undergo decays via both flavour-changing Z-boson
coupling and flavour-changing Yukawa interactions. We concentrate on the
latter channel, and study the situation where, following its pair-production,
the heavy quark pair gives rise to two tops and two Higgs boson. We show that
the case where each Higgs decays in the $b\bar{b}$ channel, there can be a
rather distinct and background-free signal that can unveil the existence of
the vectorlike isosinglet quark of this kind. 
\end{abstract}
\section{Introduction}
Our present knowledge about elementary particles and their interactions 
upto the energy scale of several hundred GeV's is encapsulated in
the theory called the Standard Model(SM). The SM, a renormalizable 
gauge theory of strong and electroweak interactions based on the guage 
group $SU(3)_{C} \times SU(2)_{L} \times U(1)_{Y}$, gives a successful 
explanation for most of the phenomena governing fundamental processes, and  
is in excellent agreement with the experimental data to date. However, there 
are a number of unanswered questions which motivates us to think beyond the SM.
These include, just to name  a few, the flavour and naturalness problems, the absence 
of a cold dark matter candidate in the spectrum, and the origin of neutrino masses and 
mixing\cite{SKM}. They have led to a plethora of conjectures  extending the SM.\\
On the experimental side, a golden opportunity to test many
of these conjectures has come through the Large Hadron Collider(LHC), 
which aims at not only discovering the only missing piece of the SM, namely the Higgs 
boson, but  also looking for new physics beyond it.
\vspace{.2cm}\\
\noindent 
One of the various ideas beyond the SM is to extend the fermionic sector of 
the SM by postulating existence of either a sequential fourth 
generation\cite{4gen} or vector-like fermions. The electroweak 
precision data strongly constrains the existence of extra chiral fermions. 
On the other hand, vector-like fermions, have left-and right-handed 
components with the same transformation property under SU(2),
and are considerably free from the aforementioned constraints. They  
can have gauge invariant mass terms of the form $\bar{\psi_L}{\psi_R}$, which
do not arise from the electroweak symmetry breaking 
mechanism, and should be traced to some new physics scale. Thus their
signature is of immediate interest if that scale is accessible to the LHC,
with the fond hope that they might contain some clue to the flavour problem.
\vspace{.2cm}\\
\noindent
Vectorlike fermions appear as singlets, doublets or triplets under  SU(2), in many 
options beyond the Standard Model(BSM)
like Little Higgs Model\cite{HCKN}, composite Higgs model\cite{ZZ}and extra 
dimensional models\cite{CDH}. They also appear in some grand unified
theories like $E_{6}$\cite{BDPW}, which once got impetus from considerations
underlying superstring theories. In particular, vectorlike 
isosinglets also play a role in the $\Delta{F}=2$ effective 
Lagrangian analysis studies\cite{ASN}. Also, it has been pointed out that
models with such an extended quark sector can
give rise to  quark electric dipole moment at the two-loop level due to the
lack of Glashow-Illiopoulos-Maiani(GIM) suppression, thereby implying
constraints on the model parameter space(s)\cite{LL}. The collider phenomenology of 
such an isosinglet vector fermion has been studied from various angles 
\cite{BL}-\cite{AFKS}.
\vspace{.2cm}\\
\noindent 
Here we focus on 
an SU(2) singlet, charge $\frac{2}{3}$ vectorlike quark. It should be noted that
down-type vectorlike isosinglets, too, have been considered extensively 
in the literature\cite{LL},\cite{AC}-\cite{AFKS}. The vectorlike quark is pair-produced 
at the LHC in the same way as the top quark, subject, 
of course, to the inevitable kinematical suppression if its mass is higher. 
Such a quark, however, has additional decay channels, which can make it 
distinct. The main features responsible for such distinction are its isosinglet 
character, and its capacity to mix with the top, once $SU(2) \times U(1)$ is 
broken. As a result of doublet-singlet mixing in the left chiral sector, such 
a quark, named \tp~ here, has flavor changing interactions both with the higgs 
boson(H) and the Z boson. We propose to utilise the resulting dacay channels, 
namely \tp $\rightarrow tH$ and \tp $\rightarrow$ tZ. In particular, we
find that the former channel leads to a rather unique and background free 
signature arising from a $t\bar{t}HH$ state.
\vspace{.2cm}\\
\noindent
On account of mixing between the SM fermions and their SU(2) singlet 
counterparts, many observables are expected to be different from the Standard 
Model predictions, specially in the sector involving third family.  
We assume a situation where Higgs is discovered already and its mass is 
approximately known. We use such a Higgs as an instrument to investigate the 
production and decay of exotic top-like quark into channels which hardly have 
Standard Model backgrounds.
We consider two Higgs masses, namely $m_H=120$ GeV and $m_H=130$ GeV. In order 
to maximize our signal rates, we let the Higgs decay in each case into the 
final state with maximum branching ratio, namely H $\rightarrow b\bar{b}$. We 
show that, using the consequent 6b final states(including decays of the top quarks 
as well), one can construct signals with very little SM backgrounds. Tagging 
five b's out of six in each case, with appropriate event selection criteria, 
proves to be sufficient for this purpose. We demonstrate that such a 
signal can allow us to probe in a discriminating fashion a large part of the 
parameter space consisting of the \tp mass and its mixing with the t.
\noindent
The rest of the paper is organised as follows. Section 2 contains an outline of the scenario, 
statements on the signal looked for,  a reminder of the constraints on various parameters,
and a resume of the methodology adopted. The results are presented in 
Section. 3. We summarise and conclude in Section 4.
\section{The scenario, the signal and the methodology}
\subsection{The scenario and its signals}
As has been said already, we consider a minimal extension of the SM with 
the inclusion of a top-ike vector isosinglet, ${t^\prime}_L$
(3,1,$+\frac{4}{3})$,~${t^{\prime}_R}$ (3,1,$+\frac{4}{3})$ to the matter 
content of the SM. The gauge boson and the Higgs sector remains unchanged.

\noindent
\tp can be produced in pair via the strong interactions or singly via 
eletroweak processes. We concentrate here on the former channel,
which at the parton level corresponds to gg $\rightarrow$ \tp\TP 
and $q\bar{q} \rightarrow$ \tp\TP~ essentially arising from the 
gluon coupling of the heavy quark:
\begin{equation}
\it{L_{QCD}}=- \iota g_s{\bar{ t^{\prime}}}\gamma_{\mu}{t^{\prime}}G_{\mu}
\end{equation}
\noindent
Neglecting the small contribution from electroweak diagrams,
the pair production cross section for an 
isosinglet fermion and a chiral fourth generation fermion 
is the same\cite{cdf},\cite{do}.The production cross section depends only upon the mass 
of \tp and goes down with increase in the mass.
\vspace{.2cm}\\
\noindent
Once the $SU(2)\times U(1)$ symmetry is broken, the most substantial mixing of 
\tp can take place with the top, as there are rather stringent bounds on mixing 
with the first two generations\cite{BB}. It must be remarked that there have 
been studies which considered the mixing with the lighter generations, which
can affect, for example, the single production of vector like quarks through 
the electroweak 
channels\cite{AACHOSU}. As a result of such mixing, the four charge $\frac{2}{3}$
 quarks in the weak basis are related to the corresponding mass eigenstates by
\begin{equation}
\begin{bmatrix}
u_0\\
c_0\\
t_0\\
t^{\prime}_0
\end{bmatrix}
={U}
\begin{bmatrix}
u\\
c\\
t\\
t^{\prime}
\end{bmatrix},
~
U=
\begin{bmatrix}
V^{\dagger}_{3 \times3} & W_{3 \times 1} \\
X_{1 \times 3} & v_{1 \times1}
\end{bmatrix}
=
\begin{bmatrix}
V_{ud}^* & V_{cd}^* & V_{td}^* & W_{dt^{\prime}}\\
V_{us}^* & V_{cs}^* & V_{ts}^* & W_{st^{\prime}}\\
V_{ub}^* & V_{cb}^* & V_{tb}^* & W_{bt^{\prime}}\\
X_{u4} & X_{c4} & X_{t4} & v_{4t^{\prime}}
\end{bmatrix}
\label{eq:U1}
\end{equation}
V is the Standard Model CKM matrix. In such a basis the mass matrix for the 
down type quarks is diagonal. The mass matrix($M^{u}$)for the up type quarks 
is
\begin{equation}
M^{u}=
\begin{bmatrix}
M_{\bar{q}_{L}q_{R}} & M_{\bar{q}_{L}t^{\prime}_R} \\
M_{\bar{t^{\prime}_{L}}{q_{R}}} & M_{\bar{t}^{\prime}_{L}t^{\prime}_R}
\end{bmatrix}
\label{eq:M1}
\end{equation}
where $q_{L,R}=(u,c,t)_{L,R}$, $M_{\bar{q}_{L}q_{R}}$ is $3 \times 3$ mass matrix of 
the SM particles, $ M_{\bar{q}_{L}t^{\prime}_R}$ is $3 \times 1$,  
$M_{\bar{q}_{L}t^{\prime}_R}$ is $1 \times 3$ and $M_{\bar{t}^{\prime}_{L}t^{\prime}_R}$ is the 
mass term for \tp. $M_{\bar{t^{\prime}_{L}}{q_{R}}}$ and $M_{\bar{t}^{\prime}_{L}t^{\prime}_R}$ do
not arise from the Yukawa couplings. $M^u$ is diagonalized by the bi-unitary 
transformation~:~$U^{\dagger}M^{u}U^{\prime}=M^{u}_{diag}$, where $U^{\prime}$ is parametrized analogously to U but in spite of right handed fermions being there in the Yukawa couplings, the elements of $U^{\prime}$ do not appear as such.\\
\noindent
With the structure of the mixing matrix being what is shown in equation(\ref{eq:U1}), 
the Standard Model CKM matrix is no longer unitary, and instead forms a block 
in the  unitary $4 \times 4$ mixing matrix U. The V and W together form the 
$4 \times 3$ charged current mixing matrix. The violation of CKM unitarity 
leads to a breakdown of the Glashow-Illiopolous-Miami(GIM) mechanism, and
leads to the flavor changing neutral current (FCNC) processes 
($t \rightarrow cZ$) and ($t \rightarrow cH$) in the top sector \cite{RR,MN}.
\noindent
As mentioned above, \tp decays to the SM fermions along with either the electroweak guage bosons 
($W^{\pm}$,Z) or Higgs boson(H) at the tree level. The charged current 
interaction (\tp$\rightarrow bW^+$) is given by
\begin{equation}
\it{L_{cc}}=\frac{gW_{bt^{\prime}}}{\sqrt2}\bar{t^{\prime}_L}\gamma_{\mu}b_LW^+_\mu
\end{equation}
On account of mixing with the top quark we have Flavor changing neutral 
current interactions with Z boson given by 
\begin{equation}
\it{L_{neutral}}=\frac{gV^*_{tb}W_{bt^{\prime}}}{2Cos_W}\bar{t^{\prime}_L}\gamma_{\mu}
t_{L}Z_{\mu} 
\end{equation}
And the interaction of \tp with Higgs boson and the Flavor changing Yukawa 
coupling is
\begin{equation}
\it{L_{yukawa}}=-y_{t^{\prime}}{\bar q}_{Li}H^{c}t^{\prime}_R + h.c
\end{equation}
where $y_{t^{\prime}}$ in this comes to be $\frac{g}{2 M_W}{V^*_{tb}}W_{b{t^{\prime}}}M_{t^{\prime}}$. In addition 
to the above yukawa coupling there are terms proportional to ${t^{\prime}_{L}}{q_R}$ and 
$\bar{t^\prime_L}{t^\prime_R}$  which arise on account of the fact that it is not possible 
to diagonalize the mass and Yukawa matrices simultaneously.\\
Following the simplified version of the mixing matrix used in \cite{MN}, we describe 
all the interactions of \tp~by the following mixing matrix
\begin{equation}
U=
\begin{bmatrix}
1 & 0 & 0 & 0\\
0 & 1 & 0 & 0\\
0 & 0 & Cos\theta & Sin\theta \\
0 & 0 & -Sin\theta & Cos\theta 
\end{bmatrix}
\label{eq:smm}
\end{equation}
Where its interactions with all quarks of the first two generations are neglected. We 
also neglect the interactions of the SM quarks across 
the generations. The elements of the mixing matrix are chosen with the motivation to 
have considerable flavor changing decay modes in addition
to the charge changing decay modes.\\
\noindent
We concentrate on one of the decay process of \tp, namely, $(t^{\prime} \rightarrow tH)$. 
For the sake of our current work we assume that the Higgs boson
has already been discovered with approximately known mass. This particular decay channel 
has already been used as a tool to study the discovery potential of Higgs\cite{AAKV,KMR,ABS} 
but with a different final state. In \cite{ABCCG} though the final decay of h 
considered is also to $b \bar{b}$
but it is in context of ``Little Higgs model'' where there are extra gauge and Higgs bosons 
along with \tp. The further decay chain we consider is where both the 
tops decay to $ bW$ and both the Higgs to $b \bar{b}$. i.e 
\begin{equation}
pp \rightarrow t^{\prime}\bar{t^{\prime}} \rightarrow hth\bar{t} \rightarrow b\bar{b}bW^+b\bar{b}\bar{b}W^-
\end{equation}
\noindent 
The final state consists of 6b's and 2 W's, out of which we attempt to identify 5b's 
and predict the signal for 5b+X final state. As we show in this work, this turns out 
to be a  rather clean signal with very low SM background.
\vspace{.2cm}\\
\noindent
Direct and indirect searches for the vector fermions put constraints on the mass
and couplings of \tp and the mixing angle $\theta$, between t and \tp. Among them, 
direct searches imply the following bounds:
\begin{itemize}
\item Considering $t^{\prime} \rightarrow W^{+}q$ as the only possible decay chain, 
the lower limit on the mass of
\tp is 358 GeV at 95$\%$ C.L. by the CDF collaboration\cite{cdf} at center of mass energy,
$\sqrt{s}$=1.96 TeV
\item The DO collaboration puts the lower limit in the channel of W+jets decay, at the 
95\% CL to be 258 GeV\cite{do} at $\sqrt{s}$=1.96 TeV.
\item The latest study by ATLAS set the lower limit on the mass of \tp at $m_{t^{\prime}} <$ 
404 GeV at the 95\% C.L. assuming 100\% decay through b$W^+$ mode\cite{atlas} at 
the center of mass energy, $\sqrt{s}$=7 TeV.
\item Assuming 100 $\%$  branching fraction for the decay \tp$\rightarrow$tZ, \tp with 
any mass less than 475 GeV is excluded at 95 $\%$ confidence level  by CMS detector 
at the LHC\cite{cms} at $\sqrt{s}$=7 TeV. 
\item The $R_b$ ratio given by, $R_B = \frac{\Gamma(Z \rightarrow b\bar{b})}{\Gamma(Z \rightarrow hadrons)}$, gives a strong constraint on the t-$t{^{\prime}}$ 
mixing angle, $\theta$ to be $\theta \le 25^{\circ}$ \cite{Vtb}.
\end{itemize}
For the purpose of our current work, we consider the lowest mass to be 350 GeV, and
make no specific assumption about branching ratios in  the three decay channels.
%
\subsection{Methodology}
For numerical calculations, we have used CalcHEP v2.5.6\cite{calc} and a CalcHEP-
PYTHIA interface program\cite{calp} along with PYTHIA-6.4.24\cite{pyth}.
The production cross-section of the isosinglet quark pair and the decay of \tp,
\TP~is calculated using CalcHEP. In order to do so, a new model with 
the new interactions based on the mixing matrix considered, was added to 
the existing list of 
CalcHEP models. We have taken $m_t$=172 GeV, and used CTEQ6L  parton 
distribution functions (PDF), for the center-of-mass energy 
$\sqrt s$=14 TeV. We chose the following benchmark points for our 
calculation:
\begin{table}[h!b]
\begin{center}
\begin{tabular}{|c|c|}
\hline
Parameter & Value \\
\hline
$M_{t^{\prime}}$[Gev] & 350,400,500\\
\hline
mixing angle $t-t^{\prime}$, $\theta$ & 5,10,15\\
\hline
$M_{h}$[GeV] & 120,130\\
\hline
\end{tabular}
\caption{Benchmark Points}
\end{center}
\label{tab:t0}
\end{table} 
After calculating the production cross section and branching fractions for all 
the benchmark points by CalcHEP, the output is transferred to the CalcHEP-PYTHIA interface 
program\cite{calp}. The interface program is provided with 
the appropriate selection cuts. Subsequently,  PYTHIA is used for obtaining 
rates in  the desired final state of 5b+X. 

\noindent
The SM background in this case is calculated at the parton level for the signal 
$pp\rightarrow hht\bar{t}$ using CalcHEP v2.5.6\cite{calc}. The only on shell 
process which gives rise to this signal in the SM is 
$pp \rightarrow t\bar{t} \rightarrow hht \bar{t}$, as a result the background 
is very small. b-identification efficiency
has been taken to be 50\%. We have  not taken into account the effect  of,
for example, charm-induced jets faking b's.  
We reconstruct the invariant mass of all the bb pairs which survive the 
first three of the selection cuts listed below, and follow the criteria explained there, 
the background for the signal 5b+X is very low in comparison to the signal.
\vspace{.2cm}\\
\noindent
We have used the following selection criteria on the minimum of five b's
required in our stipulated final state: 
\begin{itemize}
\item Each of the identified b's should have  $E^T>$40.0 GeV.
\item Each b jet should be central, with pseudorapidity, $\mid{\eta}\mid <$2.5 .
\item We implement b-tagging efficiency of 50\% i.e $\epsilon_b \sim$0.5 .
\item As a final step we calculate the invariant mass for all the possible 
combinations of b pairs. We impose the following restriction on the calculated 
invariant 
mass($m_{bb}$) i.e $m_{bb}=(M_h \pm 15)$GeV.\\ 
We get our final numbers by counting all the events(NH) which have
at least two such b pairs with their calculated invariant mass falling in the 
above limit. We predict our signal using this number.
\end{itemize}
\section{Results}
\begin{figure}[hbt]
\centering
\includegraphics[width=3.0in]{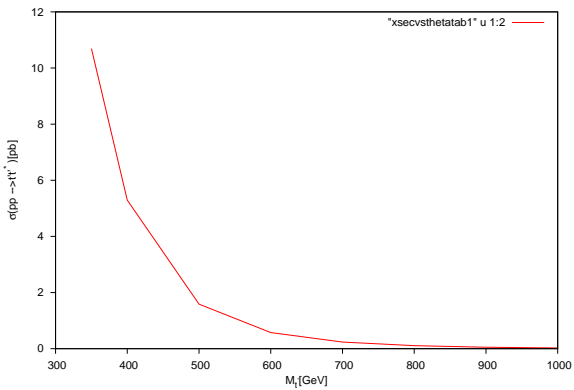}
\caption{Production Cross section of $t^{\prime}\bar{t}^{\prime}$ with the mass of $t^{\prime}$ at $\sqrt s$=14 TeV}
\label{fig:f1}
\end{figure}

\begin{figure}[h]
\centering
\includegraphics[width=3in]{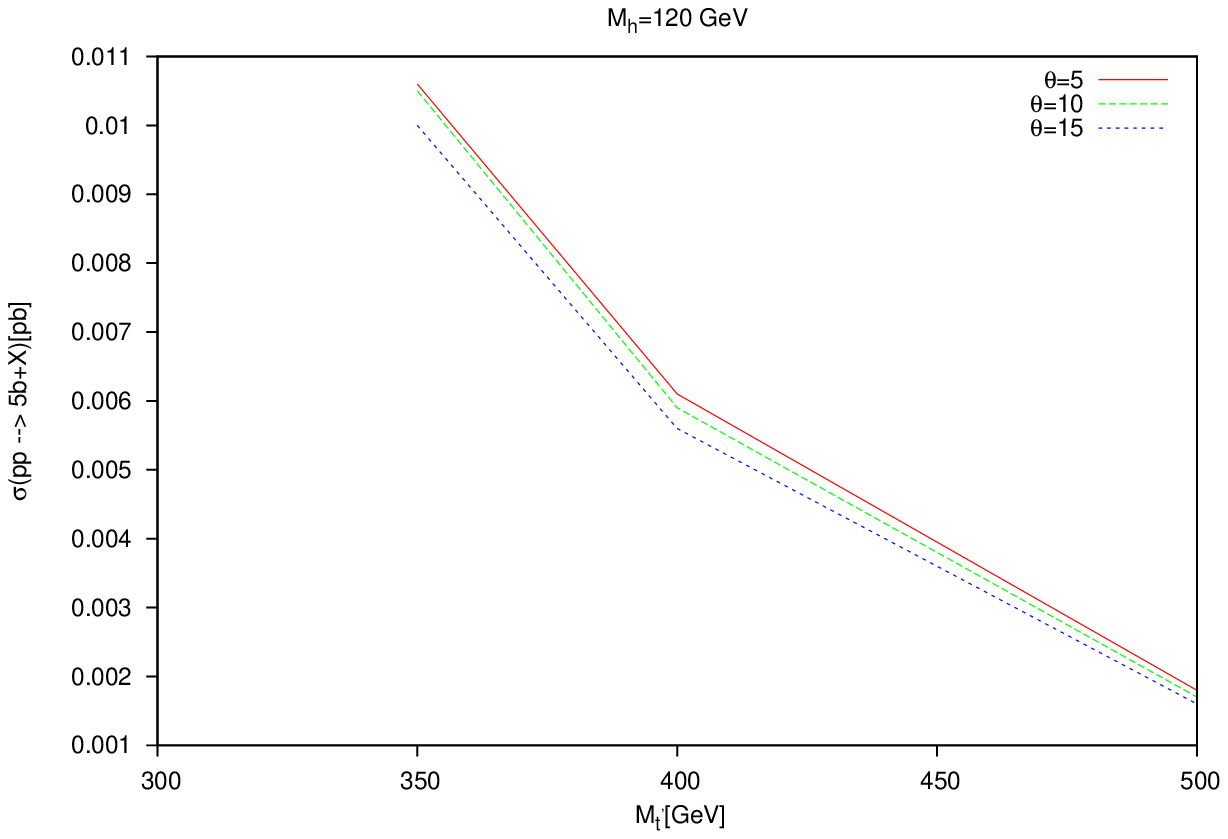}
\caption{Cross section for the final state (5b+X) with the mass of \tp for 
various values of mixing angle $\theta$ and $M_h$=120 GeV at $\sqrt s$=14 TeV}
\label{fig:f2}
\includegraphics[width=3in]{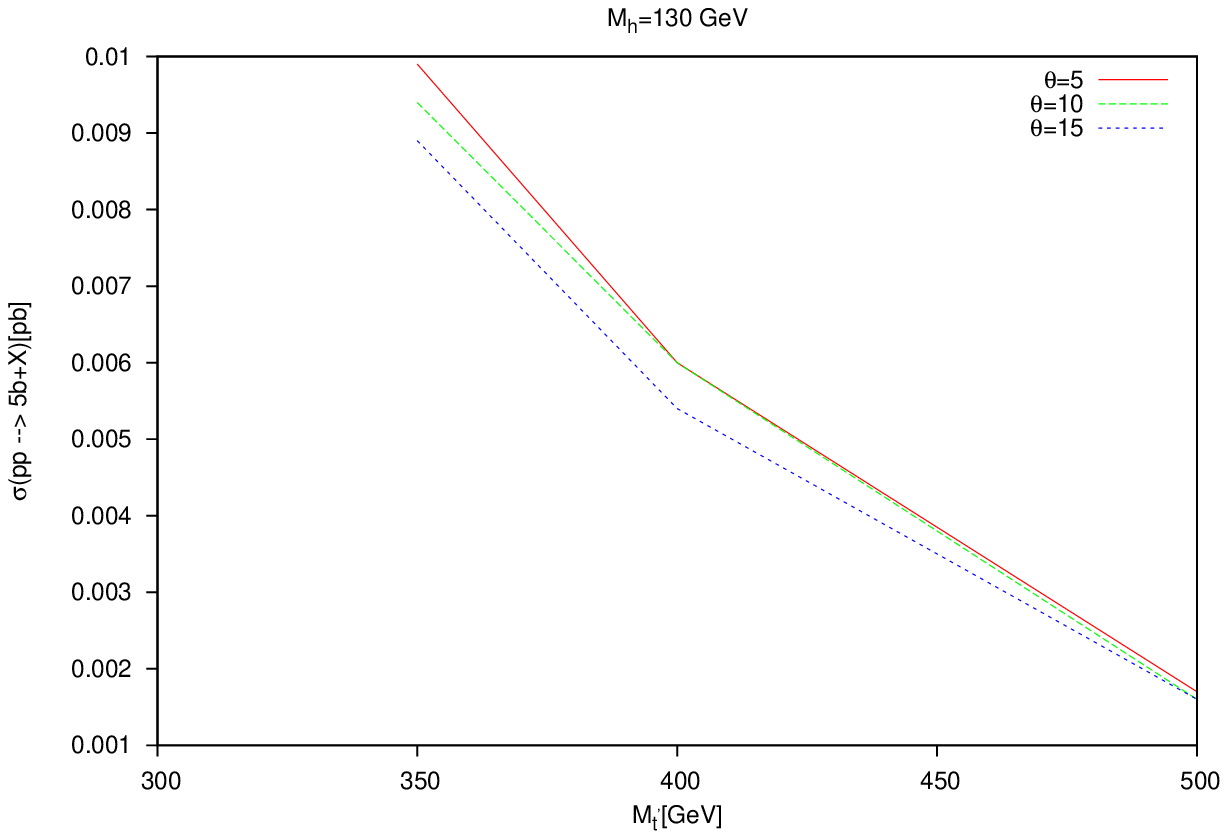}
\caption{Cross section for the final state (5b+X) with the mass of \tp for 
various values of mixing angle $\theta$ and $M_h$=130 GeV at $\sqrt s$=14 TeV}
\label{fig:f3}
\end{figure}

\begin{figure}[h]
\centering
\includegraphics[width=3in]{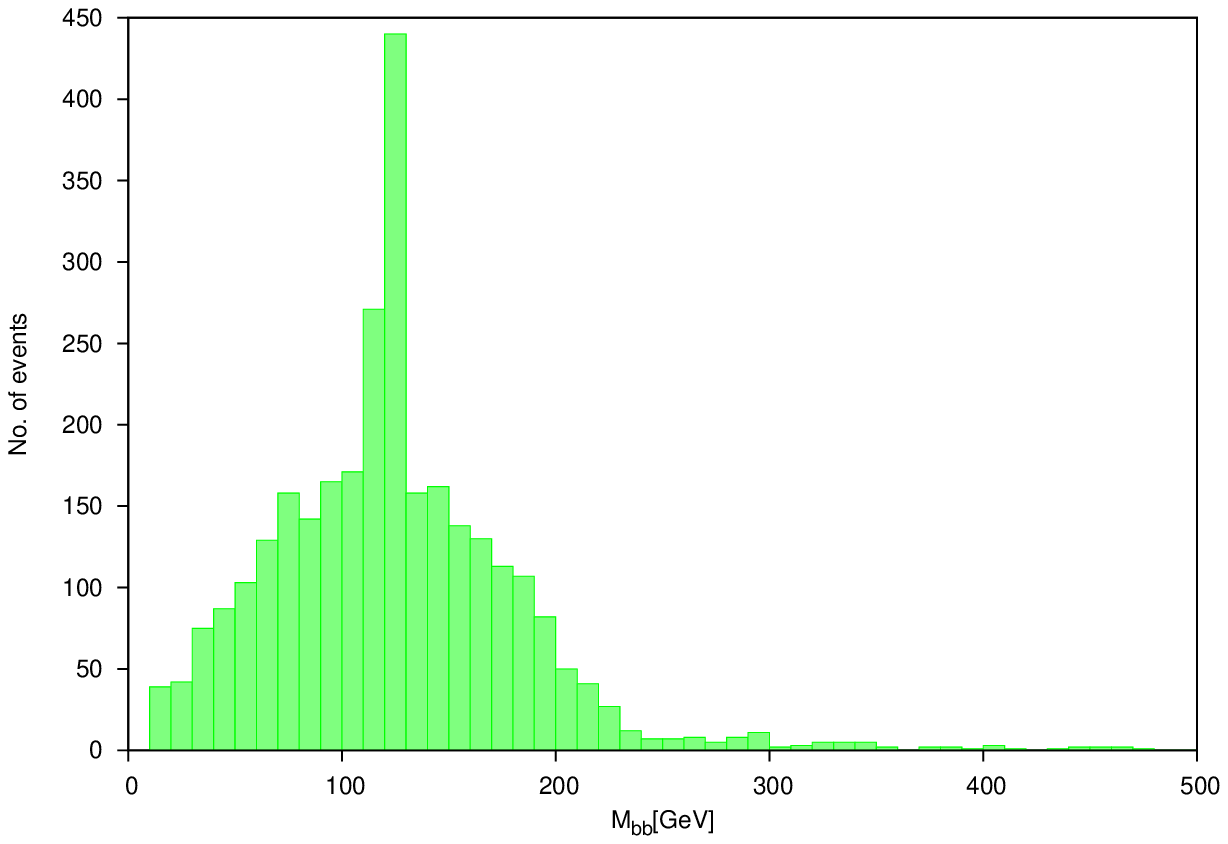}
\caption{Invariant mass distribution of $b \bar{b}$ for $m_{t^{\prime}}$=350 GeV, 
mixing angle $\theta=15$ and $M_h$=120 GeV for $\sqrt s$=14 TeV}
\label{fig:h3}
%
\includegraphics[width=3in]{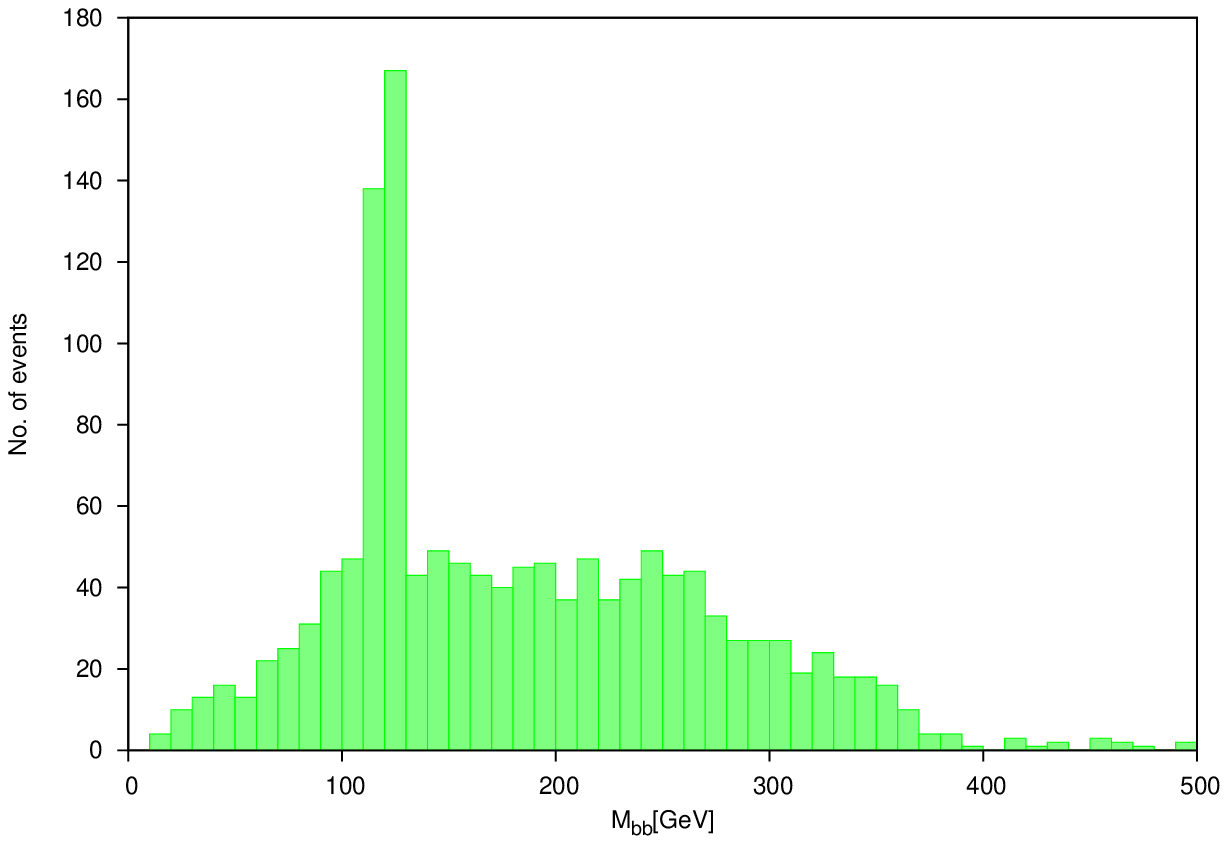}
\caption{Invariant mass distribution of $b \bar{b}$ for $m_{t^{\prime}}$=500 GeV, mixing angle $\theta=15$ and $M_h$=120 GeV for $\sqrt s$=14 TeV }
\label{fig:h1}
\end{figure}

We are presenting the results at the  leading order(LO), 
and thus our estimate can be called conservative. Our final state 
consists of six b's and two W's.
A similar use of the final state comprising six b's has been 
considered in \cite{AS}. However, the suggested signal, clean as it is, has 
suppressed rates due to the requirement of i)identification of all the six b's and ii) the simultaneous  tagging of an 
isolated lepton.
We, in contrast, give up the lepton tagging requirement. Moreover, we suggest 
identifying only five out of six b's, with the proviso that
four out of them display two individual peaks, each at the mass of the Higgs 
boson which is presumably detected before our analysis takes place. We succeed in 
considerably 
enhancing the event rates in this manner, while at the same time having 
negligible backgrounds, with our chosen acceptance criteria. As is clear 
from the plot, figure~\ref{fig:f1}, the pair production cross section of \tp  
decreases with $M_{t^{\prime}}$ since it depends only on the mass.
We find that the production of signal at the parton level, $hht\bar{t}$ in the SM, 
is weaker by atleast $10^2$ order of magnitude
in comparison to the lowest signal(for $M_t^{\prime}$=500 GeV, $M_h$=130 GeV 
and $\theta=15$) in our model as can be seen in the table (\ref{tab:t1}) 
and (\ref{tab:t2}).
\begin{table}[htpb]
\begin{center}
\begin{tabular}{|c|c|}
\hline
$M_h$(GeV) & Xsection[pb]\\
\hline 
120  & 0.00060 \\
\hline
130 & 0.00044\\
\hline
\end{tabular}
\caption{Production Cross section in the Standard Model for the signal $pp \rightarrow hht\bar{t}$ at $\sqrt s$=14 TeV }
\label{tab:t1}
\begin{tabular}{|c|c|c|c|c|}
\hline
& & \multicolumn{3}{|c|}{Cross-section[pb]}\\
\cline{3-5}
$M_{t^{\prime}}$ & $M_{h}$ [GeV] & $\theta=5$ & $\theta=10$ & $\theta=15$\\
\hline
\multirow{2}{*}{350} & 120 & 0.4272  & 0.4105 &  0.3826 \\
\cline{2-5}
& 130 & 0.3590 & 0.3583  & 0.3355\\
\hline
\hline
\multirow{2}{*}{400} & 120 & 0.2430 &  0.2365 & 0.2232\\
\cline{2-5}
& 130 & 0.2308 & 0.2236 & 0.2101\\
\hline
\hline
\multirow{2}{*}{500} & 120 & 0.07089 & 0.06931 & 0.06520 \\
\cline{2-5}
& 130 & 0.06949  &  0.06761 & 0.06324 \\
\hline
\end{tabular}
\caption{Production cross section of $pp \rightarrow t\bar{t}hh$ in our model at $\sqrt s$=14 TeV}
\label{tab:t2}
\end{center}
\end{table}
\begin{table}[h]
\begin{center}
\begin{tabular}{|c|c|c|c|c||c|c|l|}
\hline
& & \multicolumn{6}{|c|}{Cross-section[pb]}\\
\cline{3-8}
& & \multicolumn{3}{|c|}{$M_h$=120 GeV} & \multicolumn{3}{|c|}{$M_h$=130 GeV}\\
\cline{3-8}
$M_{t^{\prime}}$[GeV] & Cut & $\theta=5$ & $\theta=10$ & $\theta=15$ & $\theta=5$ & $\theta=10$ & $\theta=15$\\
\cline{1-8}
\multirow{3}{*}{350} & $E^b_T > 40.0 Gev$ & 0.1993 & 0.1907 & 
 0.1809 &  0.1730 & 0.1664 & 0.1575 \\
\cline{2-8}
& $\mid{\eta_b}\mid < 2.5$ & 0.1735 & 0.1660 &  0.1574 &  0.1519    & 0.1462 & 0.1378  \\
\cline{2-8}
& $NH \ge 2$ & .01061 & .01046 & .0098 & .0099 & .0094 & .0089  \\
\hline
\hline
\multirow{3}{*}{400} & $E^b_T > 40.0 Gev$ &  0.1155  &  0.1120  & 0.1061 & 0.1072 & 0.1038 & .0973 \\
\cline{2-8}
& $\mid{\eta_b}\mid < 2.5$ &  0.1023  & .0951 &  .0939  &  .0954 & .0925 & .0868\\
\cline{2-8}
& $NH \ge 2$ &  .00606 &  .00593 &  .00557 & .0060  & .0060  &  .0054\\
\hline
\hline
\multirow{3}{*}{500} & $E^b_T > 40.0 Gev$ & .0338 & .0323 & .0304  & .0321  & .0311 & .0292 \\
\cline{2-8}
& $\mid{\eta_b}\mid < 2.5$ & .0307  & .0293 & .0275 & .0293   & .0283 & .0267    \\
\cline{2-8}
& $NH \ge 2$ & .00175  & .00165 &  .00155 & .0017 &  .0016 & .0016    \\
\cline{1-8}
\end{tabular}
\caption{Cutflow table for various benchmark points for $M_h$=120 GeV and $M_h$=130 GeV at $\sqrt s$=14 TeV}
\label{tab:t3}
\end{center}
\end{table}
We present our results as the cut-flow chart for all the considered benchmark 
points and the invariant mass distributions of bb pairs.
We compute our results for different values of
$m_t^{\prime}$ and $M_h$ and mixing angle $\theta$. In spite of a substantial reduction of the signal due 
 to tagging efficiency of 50\%  per b, our predicted  signal is still good enough to
 be observed at the LHC at the integrated luminosity of~ 30 $fb^{-1}$. We find the following trends 
in our results as can be seen from the table \ref{tab:t3}.
\begin{itemize}
\item For a given value of $M_t^{\prime}$ and $M_h$ change in $\theta$ does not 
make much of a difference(~.2fb-.6fb)[figure \ref{fig:f2},
\ref{fig:f3}].
\item For a given value of $M_h$ and $\theta$ dependence on the $M_t^{\prime}$ is 
the strongest. It changes the cross section for the signal by 8fb-10fb.
\item For a given  $M_t^{\prime}$ and $\theta$ the change in Higgs mass does not 
really make  a difference.
\item Also as the $M_t^{\prime}$ goes from 350 GeV to 500 GeV, the change in $M_h$ 
and $\theta$ hardly makes any difference on the signal cross-section.
\end{itemize}
We plot the invariant mass distribution of all the combinations of b pairs 
for the events which survive the first three selection criteria. We find that
this distribution has two peaks corresponding to the reconstructed Higgs mass 
with in the required mass range ($m_{bb}=(M_h \pm 15)$GeV), as in figure 
\ref{fig:h3},\ref{fig:h1}, 
satisfying our fourth selection criterion, for all our benchmark points. We
 present here the distribution for two points only i.e. fig.\ref{fig:h3},
\ref{fig:h1}).
\section{Summary and Conclusions}

We have calculated the particular 
signal (5b+X), in the framework of a model with a top like($+\frac{2}{3}$) vector 
fermion, ($t^{\prime}$), in addition to the SM particles. We assume that it mixes 
only with the top quark. We ignore not only the ~interactions of \tp with first 
two generations but also the interactions of the SM quarks across the generations 
in the mixing matrix, U in equ.\ref{eq:smm}. As a result of the mixing with the top, \tp 
~has Flavor Changing Neutral interactions with Z and H bosons. We make use the 
interaction with Higgs boson and follow a particular decay mode
($t^{\prime} \rightarrow ht$) of \tp and further consider the decay of higgs to 
$b \bar{b}$. Using CalcHEP, PYTHIA and CalcHEP-PYTHIA interface programs, we 
predict an observable signal of 5b+X from a final state signal of 6b's and 
2W's, at the LHC at the center of mass energy, $\sqrt{s}$=14TeV. We consider
the situation when  the Higgs boson is already discovered, so that its mass is a 
known quantity, and can be used to identify two b-pairs with invariant mass
around the mass of the Higgs, taken here to be in the region
120 GeV-130 GeV. We find that in spite of the rather ambitious proposal of 
tagging   five b's, we get, after all cuts,  a 
signal of the order of few fb which is stronger by at least $10^2$ in comparison 
with the SM background. Since all our results are at the leading order so 
the predictions we make about the signal are rather conservative. We conclude that an 
integrated luminosity of 30 $fb^{-1}$ should be sufficient to either find out 
or rule out the existence of \tp in the mass range 350 GeV-500 GeV. 
\section{Acknowledgement} We thank Asesh Krishna Datta for
useful discussions. We acknowledge help from Sanjoy Biswas, Satyanarayan Mukhopadhaya and
Nishita Desai in carrying out the  computation. This work was partially supported by funds available
for Regional Center for Accelerator-based Particle Physics, from the
Department of Atomic Energy, Government of India.
The work of AG is supported by a grant from the Department of Science and Technology, 
Government of India, New Delhi, under the WOS-A scheme.\\ 

\noindent
{\bf{Note}}: As we were in the process of submission, a similar work(arxiv:1204.2317) appeared which considers the same decay channel of \tp and the subsequent h decay as we do but the analysis done is for different final states.

\end{document}